\documentclass[12pt,a4paper]{article}
\usepackage{amsmath}
\usepackage{bm}
\usepackage{color}
\usepackage{braket}
\usepackage{caption}
\usepackage{subcaption}
\usepackage{graphicx}
\usepackage{cite}
\usepackage{amsmath,amssymb,amsfonts}
\usepackage{algorithmic}
\usepackage{graphicx}
\usepackage{textcomp}
\usepackage{xcolor}
\usepackage{braket}
\usepackage{amsmath}
\usepackage{bm}
\usepackage{color}
\usepackage{braket}
\usepackage{caption}
\usepackage{subcaption}
\usepackage{graphicx}
\usepackage{amssymb}
\usepackage{qcircuit}

\begin{document}
\begin{center}
\begin{large}
{\bf Entanglement Of Multi-Qubit Quantum Graph States And Studies Structural Properties Of Tripartite Graphs With Quantum Computing}
\end{large}
\end{center}

\centerline {Kh. P. Gnatenko \footnote{E-Mail address: khrystyna.gnatenko@gmail.com}}
\medskip

\centerline {\small \it Ivan Franko National University of Lviv,}
\centerline {\small \it Professor Ivan Vakarchuk Department for Theoretical Physics,}
\centerline {\small \it 12 Drahomanov St., Lviv, 79005, Ukraine}

\centerline{\small \it SoftServe Inc., 2d Sadova St., 79021 Lviv, Ukraine}

\abstract{ We propose a method for constructing multi-qubit entangled quantum states that represent weighted tripartite graphs, and develop approaches for investigating their structural properties using quantum computing. In the general case of multi-qubit states corresponding to arbitrary tripartite graph structures, we derive an expression for the entanglement distance. We establish a connection between entanglement and properties of the corresponding tripartite graphs. Namely, we show that the entanglement of a qubit with the rest of the system in a quantum tripartite graph state depends on the weights of the arcs in the neighborhood of the corresponding vertex, as well as on its degree with respect to the vertex sets of the tripartite graph. As an illustrative example, we consider a tripartite graph forming a triangle and evaluate the entanglement distance with quantum computing. 
We also compute quantum correlators for the general case of tripartite quantum graph states and relate these quantities to structural features of the underlying graphs, including the number of non-overlapping neighbors, the number of common neighbors of the corresponding vertices, and the number of 4-cycles.
Obtained relationships between the quantum properties of multi-qubit states and the structural features of tripartite graphs opens up the possibility of investigating classical systems, such as tripartite graphs, using quantum computing.
It is worth emphasizing that tripartite graphs have applications in practical problems, including resource allocation, scheduling, and database and hypergraph modeling.

Keywords: entanglement distance, multi-qubit quantum graph states, tripartite graphs, quantum correlators, quantum computing
}

\section{Introduction}

At the core of quantum computing lie operations on multi-qubit entangled quantum states. Considerable attention has been devoted to the study of states that can be represented using classical objects such as graphs. These states are known as quantum graph states. In addition to their convenient graphical representation, such states are inherently entangled.
Entanglement of multi-qubit quantum states has been widely examined in the literature (see, for instance, \cite{Horodecki, Feynman, Ekert, Bennett, Lloyd, Bouwmeester, Raussendorf, Buluta, Shimony, Horodecki1, Arrigo, Shi, Llewellyn} and references therein). It serves as a fundamental resource for quantum programming and underpins many quantum algorithms and protocols. Quantum graph states have also become a prominent subject of modern investigations \cite{Bell,Vesperini1,Vesperini2, Markham,Wang,Mooney,Schlingemann, Mazurek, Shettell,  Gnatenko24, su, Hein,Guhne,Qian,Mezher}, primarily due to their wide-ranging relevance in quantum information science and quantum computing. These states describe systems of multiple qubits and admit a natural representation in terms of graphs, where vertices correspond to qubits and arcs encode their interactions. Among the studies quantum states representing bipartite graphs \cite{Gnatenko2026}, networks \cite{Gnatenko24}, and unweighted graphs \cite{Susulovska,su,Vesperini} have been examined.

Quantum graph states  play a key role in numerous quantum technologies. In particular, they appear in quantum error-correcting codes \cite{Schlingemann,Mazurek}, protocols of quantum cryptography \cite{Markham,Qian}, as well as in emerging areas such as quantum machine learning \cite{Gao,Zoufal}, among many other applications. Their versatility makes them an essential tool for both theoretical analysis and practical implementations.

In the present paper, we propose a method for constructing quantum states that can be represented by tripartite graphs. One quantitative measure of entanglement, known as the entanglement distance, was introduced in \cite{Cocc}. This measure is particularly convenient due to its direct relation to observable quantities, namely the mean spin. As a result, the entanglement can be quantified directly using quantum programming (see, for example, \cite{Gnatenko2026, GnatenkoIEEE} and references therein).

We further develop a framework for constructing quantum states associated with weighted tripartite graphs $G(U,V,W,E)$. Such graphs are characterized by the property that 
their vertices can be divided into three disjoint independent sets  $U$ $V$, $W$,  $ U \cap V = \varnothing $, $ V \cap W = \varnothing $, $ W \cap U = \varnothing $. Arcs $E$ in $G(U,V,W,E)$ are allowed only between different sets, never inside a set.
For tripartite quantum graph states corresponding to general tripartite graphs, we quantify entanglement and evaluate quantum correlators. We establish relationships between these quantum characteristics and the structural properties of tripartite graphs. In particular, the obtained results reveal connections with graph features such as the number of closed 4-cycles. These findings open up new possibilities for studying structural properties of tripartite graphs using quantum programming methods. Throughout this paper, we consider bipartite entanglement between a single qubit and the rest of the system and do not address genuine multipartite entanglement. It is worth noting that tripartite graphs have broad practical applications, including resource allocation, scheduling, and related optimization problems.

The paper is organized as follows. In Section \ref{s}, we present a method for constructing quantum states representing tripartite graphs. In Section \ref{s1}, we analytically calculate the entanglement distance for quantum graph states corresponding to tripartite graphs of arbitrary structure. In Section \ref{s2}, we evaluate quantum correlators for these states and demonstrate their relation to graph structural properties. In Section \ref{s3}, we introduce quantum protocols for studying entanglement in tripartite quantum graph states and present results of entanglement quantification obtained using the AerSimulator \cite{ibm} for specific cases. Finally, conclusions are presented in Section \ref{s4}.

\section{Multi-qubit quantum states representing  weighted tripartite graphs}\label{s}

To study properties of weighted tripartite graphs with quantum computing we propose to
consider entangled $n$-qubit quantum state 
\begin{eqnarray}
\ket{\psi}=\prod_{(u,v) \in E} \prod_{(v,w) \in E} \prod_{(u,w) \in E} RXY_{uv}(\phi_{uv})\times \nonumber\\ \times RYZ_{vw}(\phi_{vw}) RXZ_{uw}(\phi_{uw})\ket{\psi_{init}}. \label{state} 
\end{eqnarray}
The state $\ket{\psi}$ is constructed with action of two-qubit gates $RXY_{kl}(\phi_{kl})$, $RXZ_{kl}(\phi_{kl})$, $RYZ_{kl}(\phi_{kl})$ 
on the arbitrary separable quantum state 
\begin{eqnarray}
\ket{\psi_{init}}=\ket{\psi^{(U)}_{init}}\ket{\psi^{(V)}_{init}}\ket{\psi^{(W)}_{init}}.\label{state0} 
\end{eqnarray}
The two-qubit gates read
\begin{eqnarray}
RXY_{kl}(\phi_{kl})=e^{-i\frac{\phi_{kl}}{2}\sigma^x_k\sigma^y_l},\\
RXZ_{kl}(\phi_{kl})=e^{-i\frac{\phi_{kl}}{2}\sigma^x_k\sigma^z_l},\\
RYZ_{kl}(\phi_{kl})=e^{-i\frac{\phi_{kl}}{2}\sigma^y_k\sigma^z_l},
\end{eqnarray}
where  $\sigma^{j}_{k}$, $j=x,y,z$ are Pauli matrices  corresponding to qubit  $q[k]$.
For convenience, in (\ref{state0}) we use the following notations
\begin{eqnarray}
\ket{\psi^{(A)}_{init}}=\prod_{k\in A}\left(\cos \frac{\theta^{(A)}_k}{2} \ket{0}_k + e^{i\alpha^{(A)}_k} \sin \frac{\theta^{(A)}_k}{2} \ket{1}_k\right),
\end{eqnarray}
where $A$ denotes $U,V,W$.
State (\ref{state}) can be considered as a quantum graph state representing a weighted  tripartite graph $G(U,V,W,E)$.
Qubits correspond to the vertices in the graph.  The arcs  $A$ are represented with action of two-qubit gates   $RXY_{kl}(\phi_{kl})$, $RXZ_{kl}(\phi_{kl})$, $RYZ_{kl}(\phi_{kl})$. Parameters of the gates  $\phi_{kl}$  characterize weights of  arcs 
$\{k, l\}$ in  $G(U,V,W,E)$. Arcs linked vertices in sets $U$, $V$ are represented with action of $RXY_{uv}(\phi_{uv})$  gates on the initial state of qubits $q[u]$, $q[v]$, $u \in U$, $v \in V$. Action of gates $RXY_{uv}(\phi_{uv})$, $RYX_{vu}(\phi_{vu})$    represents  arcs $\{u, v\}$  with weights $\phi_{uv}$,  $u \in U$, $v \in V$, respectively.
Similarly, for arcs linking vertices in sets $V$, $W$ and  $U$, $W$ we consider gates $RYZ_{vw}(\phi_{vw})$, $RXZ_{uw}(\phi_{uw})$. 
Note that for a tripartite graph, the arc link vertices within different sets $U$, $V$, $W$.   There are no adjacent vertices (vertices linked with arc) in the same set. Therefore the corresponding gates $RXY_{uv}(\phi_{uv})$, $RXZ_{uw}(\phi_{uw})$, $RYZ_{vw}(\phi_{vw})$  commute with each other.  

So, in this paper, we introduce a class of multi-qubit quantum states associated with weighted tripartite graphs. These states differ from the standard graph states as they are generated by a different set of two-qubit interactions acting on a general separable $n$-qubit state, where $n$ is equal to the number of vertices $V$ in the graph $G(U,V,W,E)$. The three types of two-qubit rotation gates, $RXY$, $RYZ$, $RZX$, are introduced to represent the three different classes of edges in a weighted tripartite graph.

It is worth noting that the states considered in this work differ from the well-studied graph states, which are constructed by applying $CZ$ gates to the $|+\rangle^{\otimes n}$ state (see, for instance, \cite{Markham, Wang, Mooney} and references therein). Throughout this paper,  we adopt the general notion of a quantum graph state as any quantum state that can be uniquely represented by a graph \cite{Gnatenko2026,Susulovska}.

\section{Entanglement distance of  multi-qubit states  corresponding to general tripartite graphs}\label{s1}

Let us find the entanglement  of qubits in multi-qubit quantum graph states corresponding to general tripartite graphs. The entanglement distance of qubit $q[k]$ with other qubits in quantum state $\ket{\psi}$, $E^{ED}_k (\ket{\psi})$ is related to the mean value of spin as follows
\begin{eqnarray}
E^{ED}_k (\ket{\psi}) = 1-\sum_{j=x,y,z}\bra{\psi}\sigma^{j}_{k}\ket{\psi}^2, \label{ed}
\end{eqnarray}
where $E^{ED}_k (\ket{\psi})$ is the entanglement of $q[k]$ with other qubits in quantum state $\ket{\psi}$,  \cite{Vesperini, Cocc}.

To find entanglement of qubits representing vertices in different sets $U$, $V$, $W$ we calculate mean values $\bra{\psi}\sigma^{\alpha}_u\ket{\psi}$, $\bra{\psi}\sigma^{\alpha}_v\ket{\psi}$, $\bra{\psi}\sigma^{\alpha}_w\ket{\psi}$, $\alpha=x,y,z$ in general case of tripartite quantum state $\ket{\psi}$ (\ref{state}).

For $\bra{\psi} \sigma^x_u \ket{\psi}$, $\bra{\psi} \sigma^y_v \ket{\psi}$, $\bra{\psi} \sigma^z_w \ket{\psi}$,  $u \in U$, $v \in V$, $w \in W$ we have 
\begin{eqnarray}
\bra{\psi} \sigma^x_u \ket{\psi} =
\bra{\psi_{init}}\sigma^x_u \ket{\psi_{init}}=
\cos \alpha^{(U)}_u \sin \theta^{(U)}_u,\\
\bra{\psi} \sigma^y_v \ket{\psi} =
\bra{\psi_{init}}\sigma^y_v \ket{\psi_{init}}=
\sin \alpha^{(V)}_v \sin \theta^{(V)}_v, \label{yvv}\\
\bra{\psi} \sigma^z_w \ket{\psi} =
\bra{\psi_{init}}\sigma^z_w \ket{\psi_{init}}=\cos\theta^{(W)}_w.\label{yvv1}
\end{eqnarray}

Let us calculate also the mean values $\bra{\psi} \sigma^y_u \ket{\psi}$, $\bra{\psi} \sigma^z_u \ket{\psi}$, $u\in U$
one finds
\begin{eqnarray}
\bra{\psi} \sigma^y_u \ket{\psi} =\nonumber\\
\bra{\psi_{init}}\prod_{u^{\prime},v^{\prime} \in E}\prod_{u^{\prime},w^{\prime} \in E}  RXY^{+}_{u^{\prime}v^{\prime}}(\phi_{u^{\prime}v^{\prime}})\times \nonumber\\ \times RXZ^{+}_{u^{\prime}w^{\prime}}(\phi_{u^{\prime}w^{\prime}}) \sigma^y_u \times \nonumber\\ \times \prod_{u^{\prime\prime},v^{\prime\prime} \in E}\prod_{u^{\prime\prime},w^{\prime\prime} \in E}  RXY_{u^{\prime\prime}v^{\prime\prime}}(\phi_{u^{\prime\prime}v^{\prime\prime}})\times \nonumber\\ \times  RXZ_{u^{\prime\prime}w^{\prime\prime}}(\phi_{u^{\prime\prime}w^{\prime\prime}})  \ket{\psi_{init}}=
\nonumber\\
\bra{\psi_{init}} \prod_{v \in N_{V} (u)} \prod_{w \in N_{W}(u)}
RXY^{+}_{uv}(2\phi_{uv})\times \nonumber\\ \times RXZ^{+}_{uw}(2\phi_{uw})\sigma^y_u  \ket{\psi_{init}}=\Re{a_u},
 \label{uy}
 \end{eqnarray}
 \begin{eqnarray}
 \bra{\psi} \sigma^z_u \ket{\psi} =
\nonumber\\
\bra{\psi_{init}} \prod_{v \in N_{V} (u)} \prod_{w \in N_{W}(u)}
RXY^{+}_{uv}(2\phi_{uv})\times \nonumber\\ \times RXZ^{+}_{uw}(2\phi_{uw})\sigma^z_u  \ket{\psi_{init}}=\Im{a_u}.\label{uz} 
\end{eqnarray}
The complex number $a_u$ in (\ref{uy}), (\ref{uz}) is as follows
\begin{eqnarray}
a_u=\left(\sin \alpha_u \sin \theta_u + i \cos \theta_u \right) \times \nonumber\\
\mathop{\prod_{v \in N_{V}(u)}}\mathop{\prod_{w \in N_{W}(u) }} \left(\cos \phi_{uv} + i \sin \phi_{uv} \sin \alpha_v \sin \theta_v\right) \times \nonumber\\ \times \left(\cos \phi_{uw} + i \sin \phi_{uw}  \cos \theta_w\right). 
\end{eqnarray}
Here $N_{V}(u) = \{ v \in V \mid (u,v) \in E \}$, $N_{W}(u) = \{ w \in W \mid (u,w) \in E \}$ are set of vertices in $V$  adjacent to $u$ and set of vertices in $W$  adjacent to $u$, respectively.  

So, for arbitrary structure of tripartite graph the entanglement of qubit $q[u]$ ($u\in U$) with other qubits in state (\ref{state}) reads

\begin{eqnarray}
E^{ED}_u(\ket{\psi}) = \left(\sin^2 \alpha_u \sin^2 \theta_u + \cos^2 \theta_u\right)(1- \nonumber\\ \mathop{\prod_{v \in N_{V(u)}}\mathop{\prod_{w \in N_{W}(u) }}} \left(\cos^2 \phi_{uv} +  \sin^2 \phi_{uv} \sin^2\alpha_v \sin^2 \theta_v\right) \times \nonumber\\ \times \left(\cos^2 \phi_{uw} + \sin^2 \phi_{uw}  \sin^2 \theta_w\right)). \label{edu}
\end{eqnarray}
So, the entanglement $E^{ED}_u(\ket{\psi})$ is determined by the weights of the arcs incident to the vertex $u$, described by the parameters $\phi_{uv}$ and $\phi_{uw}$ and also it depends on the parameters of initial states of qubits representing close
neighborhood of $u$ in  $G(U,V,W,E)$.

Similarly for $\bra{\psi} \sigma^x_v \ket{\psi}$, $\bra{\psi} \sigma^z_v \ket{\psi}$, $v\in V$ one finds

\begin{eqnarray}
\bra{\psi} \sigma^x_v \ket{\psi} =\nonumber\\
\bra{\psi_{init}} \prod_{u \in N_{U} (v)} \prod_{w \in N_{W}(v)}
RYZ^{+}_{vw}(2\phi_{vw})\times \nonumber\\ \times RXY^{+}_{uv}(2\phi_{vu})\sigma^x_v  \ket{\psi_{init}}=\Im{c_v}
 \label{vx}\\
 \bra{\psi} \sigma^z_v \ket{\psi} =\Re{c_v} \label{vz},\nonumber\\
c_v=\left(\cos \alpha_v \sin \theta_v + i \cos \theta_v \right) \times \nonumber\\
 \prod_{u \in N_{U} (v)} \prod_{w \in N_{W}(v)} \left(\cos \phi_{uv} + i \sin \phi_{uv} \cos \alpha_u \sin \theta_u\right)\times \nonumber\\ \times  \left(\cos \phi_{vw} + i \sin \phi_{vw}  \cos \theta_w\right). 
\end{eqnarray}

Using these results and taking equation (\ref{yvv}) into account, we can express the entanglement between $q[v]$ (with $v \in V$) and the remaining qubits in the arbitrary graph state (\ref{state}) as follows:

\begin{eqnarray}
E^{ED}_v (\ket{\psi}) = 1-\sin^2 \alpha_v\sin^2 \theta_v-\nonumber\\-\left(\cos^2 \alpha_v \sin^2 \theta_v + \cos^2 \theta_v\right) \times \nonumber\\
 \prod_{u \in N_{U} (v)} \prod_{w \in N_{W}(v)} \left(\cos^2 \phi_{uv} +  \sin^2 \phi_{uv} \cos^2\alpha_v \sin^2 \theta_v\right) \times \nonumber\\ \times \left(\cos^2 \phi_{vw} + \sin^2 \phi_{vw}  \cos^2 \theta_w\right) . \label{edu1}
\end{eqnarray}

As a result, we conclude the entanglement $E^{ED}_v(\ket{\psi})$ is governed by the edge weights connected to the vertex $v$, characterized by the parameters $\phi_{uv}$ and $\phi_{vw}$, as well as by the parameters defining the initial states of the qubits in the vicinity of $v$ within the graph $G(U,V,W,E)$.

In addition, for $\bra{\psi} \sigma^x_w \ket{\psi}$, $\bra{\psi} \sigma^z_w \ket{\psi}$, $w\in W$ one finds

\begin{eqnarray}
\bra{\psi} \sigma^x_w \ket{\psi} =\nonumber\\
\bra{\psi_{init}} \prod_{u \in N_{U} (w)} \prod_{v \in N_{V}(w)}
RXZ^{+}_{uw}(2\phi_{uw})\times \nonumber\\ \times RYZ^{+}_{vw}(2\phi_{vw})\sigma^x_w  \ket{\psi_{init}}=2\Re{c_w},
 \label{wx}\\
 \bra{\psi} \sigma^y_w \ket{\psi} =2\Im{c_w} \label{vz},\nonumber\\
c_w= e^{\alpha_w}\sin \theta_w \times \nonumber\\
 \prod_{u \in N_{U} (w)} \prod_{v \in N_{V}(w)} \left(\cos \phi_{uw} + i \sin \phi_{uw} \cos \alpha_u \sin \theta_u\right)\times \nonumber\\ \times  \left(\cos \phi_{vw} + i \sin \phi_{vw} \sin \alpha_v  \sin \theta_v\right). 
\end{eqnarray}

Building on these results and incorporating Eq.~(\ref{yvv1}), we obtain an expression for the entanglement between $q[w]$ (with $w \in W$) and the remaining qubits in the arbitrary graph state (\ref{state}) as follows:

\begin{eqnarray}
E^{ED}_w (\ket{\psi}) = \sin^2 \theta_w(1- \nonumber\\
\prod_{u \in N_{U} (w)} \prod_{v \in N_{V}(w)} \left(\cos^2 \phi_{vw} +  \sin^2 \phi_{vw} \sin^2\alpha_v \sin^2 \theta_v\right) \times \nonumber\\ \times \left(\cos^2 \phi_{uw} + \sin^2 \phi_{uw}  \cos^2 \alpha_u\sin^2 \theta_u\right)) . \label{edu2}
\end{eqnarray}

Note that, in the particular case of an unweighted tripartite graph $G(U,V,W,E)$ with $\phi_{ij}=\phi$, and for identical parameters in the initial state $\ket{\phi_{init}}$, i.e., $\alpha_i=\alpha$ and $\theta_i=\theta$, taking into account the obtained results (\ref{edu}), (\ref{edu1}), and (\ref{edu2}), we have

\begin{eqnarray}
E^{ED}_u (\ket{\psi}) = \left(\sin^2 \alpha\sin^2 \theta + \cos^2 \theta\right)(1-  \nonumber\\
\left(\cos^2 \phi+  \sin^2 \phi\sin^2\alpha \sin^2 \theta\right)^{\vert N_{V}(u)\vert}  \times \nonumber\\ \times \left(\cos^2 \phi + \sin^2 \phi  \sin^2 \theta \right)^{\vert N_{W}(u)\vert}), \label{edu}
\end{eqnarray}

\begin{eqnarray}
E^{ED}_v (\ket{\psi}) = 1-\sin^2 \alpha\sin^2 \theta-\nonumber\\-\left(\cos^2 \alpha \sin^2 \theta + \cos^2 \theta\right) \times \nonumber\\
\left(\cos^2 \phi +  \sin^2 \phi \cos^2\alpha \sin^2 \theta \right)^{\vert N_{U}(v)\vert} \times \nonumber\\ \times \left(\cos^2 \phi+ \sin^2 \phi  \cos^2 \theta\right)^{\vert N_{W}(v)\vert},\label{cedu1}
\end{eqnarray}

\begin{eqnarray}
E^{ED}_w (\ket{\psi}) = \sin^2 \theta(1- \nonumber\\
\left(\cos^2 \phi+  \sin^2 \phi\sin^2\alpha \sin^2 \theta\right)^{\vert N_{V}(w)\vert} \times \nonumber\\ \times \left(\cos^2 \phi+ \sin^2 \phi \cos^2 \alpha\sin^2 \theta\right)^{\vert N_{U}(w)\vert}.\label{cedu2}
\end{eqnarray}

Here, $\lvert ...\rvert$ denotes the cardinality of a set.
Thus, the entanglement of a given qubit with the rest of the system in a quantum state corresponding to an unweighted tripartite graph depends on the degree of the corresponding vertex with respect to other sets. Namely, the entanglement of the qubit $q[u]$ (with $u \in U$) with the remaining qubits in the arbitrary graph state (\ref{edu}) depends on the degrees of the vertex $u$ with respect to the subsets $V$ and $W$, namely $|N_{V}(u)|$ and $|N_{W}(u)|$.Similarly, the entanglement of the qubit $q[v]$ (with $v \in V$) with the other qubits in (\ref{cedu1}) is determined by the degrees of the vertex $v$ with respect to the subsets $U$ and $W$, namely $|N_{U}(v)|$ and $|N_{W}(v)|$. Finally, the entanglement of the qubit $q[w]$ (with $w \in W$) with the remaining qubits in (\ref{cedu2}) depends on the degrees of the vertex $w$ with respect to the subsets $U$ and $V$, namely $|N_{U}(w)|$ and $|N_{V}(w)|$. So, the quantum property as entanglement is related with the degrees of corresponding vertices in graph.

In the next section, we present calculations of quantum correlators and their dependence on the structural properties of tripartite graphs.

\section{Relation of quantum correlators with structural properties of tripartite graphs}\label{s2}

In this section, we calculate quantum correlators in multi-qubit quantum states representing tripartite graphs (\ref{state}). We show that these  values are related to properties of the corresponding graphs. 

Result for $\bra{\psi} \sigma^x_{u_1}  \sigma^x_{u_2}  \ket{\psi}$ with ${u_1} \in U$, ${u_2}\in U$
reads
\begin{eqnarray}
\bra{\psi} \sigma^x_{u_1}  \sigma^x_{u_2}  \ket{\psi}=\sin\theta_{u_1}\sin\theta_{u_2}\cos\alpha_{u_1}\cos\alpha_{u_2}.
\end{eqnarray}
Let us calculate 
$\bra{\psi} \sigma^y_{u_1}  \sigma^y_{u_2}  \ket{\psi}$, 
$\bra{\psi} \sigma^z_{u_1}  \sigma^z_{u_2}  \ket{\psi}$ with ${u_1} \in U$, ${u_2}\in U$, 
 $u_1\neq u_2$. 
We can write
\begin{eqnarray}
\bra{\psi} \sigma^y_{u_1}  \sigma^y_{u_2}  \ket{\psi}
=
 \bra{\psi_{init}} \prod_{v_1\in N_{V}(u_1)} \prod_{v_2\in N_{V}(u_2)} \nonumber\\ \prod_{w_1\in N_{W}(u_1)} \prod_{w_2\in N_{W}(u_2)}RXY^{+}_{{u_1} {v_1}} (2\phi_{{u_1} {v_1}})\times \nonumber\\ \times RXY^{+}_{{u_2} {v_2}}(2\phi_{{u_2} {v_2}})RXZ^{+}_{{u_1} {w_1}}(2\phi_{{u_1} {w_1}}) \times \nonumber\\ \times  RXZ^{+}_{{u_2} {w_2}}(2\phi_{{u_2} {w_2}})  \sigma^y_{u_1}   \sigma^y_{u_2}   \ket{\psi_{init}}=\nonumber\\= \frac{1}{2} \Re(-z^{(1)}_{{u_1}{u_2}} a_{u_1}a_{u_2}+z^{(2)}_{{u_1}{u_2}}a_{u_1}a^*_{u_2}), \label{xxu}
\end{eqnarray}

\begin{eqnarray}
\bra{\psi} \sigma^z_{u_1}  \sigma^z_{u_2}  \ket{\psi}=
 \bra{\psi_{init}} \prod_{v_1\in N_{V}(u_1)} \prod_{v_2\in N_{V}(u_2)} \nonumber\\\prod_{w_1\in N_{W}(u_1)} \prod_{w_2\in N_{W}(u_2)}RXY^{+}_{{u_1} {v_1}} (2\phi_{{u_1} {v_1}})\times \nonumber\\ \times RXY^{+}_{{u_2} {v_2}}(2\phi_{{u_2} {v_2}}) RXZ^{+}_{{u_1} {w_1}}\times \nonumber\\ \times(2\phi_{{u_1} {w_1}}) RXZ^{+}_{{u_2} {w_2}}(2\phi_{{u_2} {w_2}}) \sigma^z_{u_1}   \sigma^z_{u_2}   \ket{\psi_{init}}=\nonumber\\=\frac{1}{2} \Re (z^{(1)}_{{u_1}{u_2}} a_{u_1}a_{u_2}+z^{(2)}_{{u_1}{u_2}}a_{u_1}a^*_{u_2}),\label{zzu}
\end{eqnarray}

For convenience in (\ref{xxu}), (\ref{zzu}) we introduced the following notations
\begin{eqnarray}
a_{u}=\cos\theta_{u}-i\sin\alpha_{u}\sin\theta_{u}, \label{a}
\end{eqnarray}
\begin{eqnarray}
z^{(1)}_{{u_1}{u_2}} =\nonumber\\ \prod_{v_1\in N_{V}(u_1)} \prod_{\substack{v_2\in N_{V}(u_2)\\ v_2 \neq v_1}}(\cos \phi_{u_1v_1} +i\sin\theta_{v_1}\sin\alpha_{v_1}\sin \phi_{u_1v_1} )\times\nonumber\\ \times (\cos \phi_{u_2v_2} +i\sin\theta_{v_2}\sin\alpha_{v_2}\sin \phi_{u_2v_2} )\times\nonumber
\end{eqnarray}
\begin{eqnarray}
\prod_{w_1\in N_{W}(u_1)} \prod_{\substack{w_2\in N_{W}(u_2)\\ w_2 \neq w_1}} (\cos \phi_{u_1w_1} +i\cos\theta_{w_1}\sin \phi_{u_1w_1} )\times \nonumber\\ \times (\cos \phi_{u_2w_2} +i\cos\theta_{w_2}\sin \phi_{u_2w_2} )\times\nonumber\\ 
\prod_{v_3\in N_{V}(u_1,u_2)} (\cos(\phi_{u_1v_3}+\phi_{u_2v_3}) +\nonumber\\+i\sin\theta_{v_3}\sin\alpha_{v_3}\sin (\phi_{u_1v_3}+\phi_{u_2v_3}) )\nonumber\\ \prod_{w_3\in N_{W}(u_1,u_2)}(\cos(\phi_{u_1w_3}+\phi_{u_2w_3}) +\nonumber\\+i\cos\theta_{w_3}\sin (\phi_{u_1w_3}+\phi_{u_2w_3}) ),\nonumber\\
\end{eqnarray}
\begin{eqnarray}
z^{(2)}_{{u_1}{u_2}} =\nonumber\\\prod_{v_1\in N_{V}(u_1)} \prod_{\substack{v_2\in N_{V}(u_2)\\ v_2 \neq v_1}}(\cos \phi_{u_1v_1} +i\sin\theta_{v_1}\sin\alpha_{v_1}\sin \phi_{u_1v_1} )\times\nonumber\\ \times (\cos \phi_{u_2v_2} -i\sin\theta_{v_2}\sin\alpha_{v_2}\sin \phi_{u_2v_2} )\times\nonumber\\ 
\prod_{w_1\in N_{W}(u_1)} \prod_{\substack{w_2\in N_{W}(u_2)\\ w_2 \neq w_1}} (\cos \phi_{u_1w_1} +i\cos\theta_{w_1}\sin \phi_{u_1w_1} )\times  \nonumber
\end{eqnarray}

\begin{eqnarray}
\times (\cos \phi_{u_2w_2} -i\cos\theta_{w_2}\sin \phi_{u_2w_2} )\times\nonumber\\ \times
\prod_{v_3\in N_{V}(u_1,u_2)} (\cos(\phi_{u_1v_3}-\phi_{u_2v_3}) +\nonumber\\+i\sin\theta_{v_3}\sin\alpha_{v_3}\sin (\phi_{u_1v_3}-\phi_{u_2v_3}) )\times\nonumber\\ \prod_{w_3\in N_{W}(u_1,u_2)}(\cos(\phi_{u_1w_3}-\phi_{u_2w_3}) +\nonumber\\+i\cos\theta_{w_3}\sin (\phi_{u_1w_3}-\phi_{u_2w_3}) ),
\end{eqnarray}
here $N_{A}(a)$ is the set of  neighbors of vertex $a$ in $A$,  and 
$N_{A}(u_1,u_2)$ in a set of common neighbors of vertexes $u_1$, $u_2$ in set $A$, ($A$ denotes $V,W$ in tripartite graph $G(U,V,W,E)$).  Also, for $\bra{\psi} \sigma^y_{u_1}  \sigma^z_{u_2}  \ket{\psi}$ we obtain

\begin{eqnarray}
\bra{\psi} \sigma^y_{u_1}  \sigma^z_{u_2}  \ket{\psi}=-\frac{1}{2} \Im (z^{(1)}_{{u_1}{u_2}} a_{u_1}a_{u_2}+z^{(2)}_{{u_1}{u_2}}a_{u_1}a^*_{u_2}), \label{yzu}
\end{eqnarray}

In the case of uniform parameters for the gates $RXY_{uv}(\phi_{uv})$, $RXZ_{uw}(\phi_{uw})$, and $RYZ_{vw}(\phi_{vw})$, namely $\phi_{uv}=\phi_{uw}=\phi_{vw}=\phi$, and assuming that all qubits within each set are initialized in identical states,
 $\theta_{v_i}=\theta^{(V)}$, $\alpha_{v_i}=\alpha^{(V)}$,  
$\theta_{w_i}=\theta^{(W)}$, $\alpha_{w_i}=\alpha^{(W)}$, we can write
%%%%%%%%%%%%%%%%%%%%%%%%%%%%%%%%%%

\begin{eqnarray}
z^{(1)}_{{u_1}{u_2}} =\nonumber\\=(\cos \phi +i\sin\theta^{(V)}\sin\alpha^{(V)}\sin \phi )^{\lvert N_V(u_1) \,\triangle\, N_V(u_2) \rvert}\times 
\nonumber\\ \times
(\cos \phi +i\cos\theta^{(W)}\sin \phi )^{\lvert N_W(u_1) \,\triangle\, N_W(u_2) \rvert}\times\nonumber\\ \times
(\cos(2\phi) +i\sin\theta^{(V)}\sin\alpha^{(V)}\sin (2\phi) )^{\vert N_V(u_1) \cap N_V(u_2)\vert}\nonumber\\ (\cos(2\phi) +i\cos\theta^{(W)}\sin (2\phi) )^{\vert N_W(u_1) \cap N_W(u_2)\vert}
\end{eqnarray}
\begin{eqnarray}
z^{(2)}_{{u_1}{u_2}} =\nonumber\\=(\cos \phi +i\sin\theta^{(V)}\sin\alpha^{(V)}\sin \phi )^{\lvert N_V(u_1) / N_V(u_2) \rvert}\times \nonumber\\ \times (\cos \phi -i\sin\theta^{(V)}\sin\alpha^{(V)}\sin \phi )^{\lvert N_V(u_2) / N_V(u_1) \rvert}\times 
\nonumber\\ \times
(\cos \phi +i\cos\theta^{(W)}\sin \phi )^{\lvert N_W(u_1)/N_W(u_2) \rvert}\times \nonumber\\ \times (\cos \phi -i\cos\theta^{(W)}\sin \phi )^{\lvert N_W(u_2)/N_W(u_1) \rvert}.
\end{eqnarray}
Here $\lvert N_A(u_1) \,\triangle\, N_A(u_2) \rvert$
is the cardinality of  the symmetric difference of the sets $N_A(u_1)$, $N_A(u_2)$ which corresponds to the number of different (non-overlapping) neighbors of $u_1$, $u_2$; and $\lvert N_A(u_1) \,\cap\, N_A(u_2) \rvert$
is equal to the number of common neighbors of $u_1$, $u_2$ in $A$,  ($A$ denotes $V,W$ in tripartite graph $G(U,V,W,E)$).  It is important to note that the value $\lvert N_A(u_1) \cap N_A(u_2) \rvert$ determines the number of 4-cycles involving the vertices $u_1$, $u_2$, and vertices from the set $A$ in the tripartite graph. Namely the number reads
\begin{eqnarray}
n^{(A)}_4=C^2_{\lvert N_A(u_1) \,\cap\, N_A(u_2) \rvert}.
\end{eqnarray}

For $\bra{\psi} \sigma^{x}_{v_1}  \sigma^x_{v_2}  \ket{\psi}$, $\bra{\psi} \sigma^y_{v_1}  \sigma^y_{v_2}  \ket{\psi}$, $\bra{\psi} \sigma^z_{v_1}  \sigma^z_{v_2}  \ket{\psi}$, $\bra{\psi} \sigma^z_{v_1}  \sigma^x_{v_2}  \ket{\psi}$  with ${v_1} \in V$, ${v_2}\in V$, 
 $v_1\neq v_2$  we find
\begin{eqnarray}
\bra{\psi} \sigma^x_{v_1}  \sigma^x_{v_2}  \ket{\psi}=
 \bra{\psi_{init}} \prod_{u_1\in N_{U}(v_1)} \prod_{u_2\in N_{U}(v_2)} \nonumber\\\prod_{w_1\in N_{W}(v_1)} \prod_{w_2\in N_{W}(v_2)}RXY^{+}_{{u_1} {v_1}} (2\phi_{{u_1} {v_1}})\times \nonumber\\ \times RXY^{+}_{{u_2} {v_2}}(2\phi_{{u_2} {v_2}})\times \nonumber\\ \times RYZ^{+}_{{v_1} {w_1}}(2\phi_{{v_1} {w_1}}) RYZ^{+}_{{v_2} {w_2}}(2\phi_{{v_2} {w_2}}) \sigma^x_{v_1}   \sigma^x_{v_2}   \ket{\psi_{init}}=\nonumber\\=\frac{1}{2} \Re (-z^{(1)}_{{v_1}{v_2}} a_{v_1}a_{v_2}+z^{(2)}_{{v_1}{v_2}}a_{v_1}a^*_{v_2}),
 \end{eqnarray}
 \begin{eqnarray}
\bra{\psi} \sigma^z_{v_1}  \sigma^z_{v_2}  \ket{\psi}=\frac{1}{2} \Re (z^{(1)}_{{v_1}{v_2}} a_{v_1}a_{v_2}+z^{(2)}_{{v_1}{v_2}}a_{v_1}a^*_{v_2}),
 \end{eqnarray}
 \begin{eqnarray}
\bra{\psi} \sigma^z_{v_1}  \sigma^x_{v_2}  \ket{\psi}=\frac{1}{2} \Im (z^{(1)}_{{v_1}{v_2}} a_{v_1}a_{v_2}+z^{(2)}_{{v_1}{v_2}}a_{v_1}a^*_{v_2}),
\end{eqnarray}
here $a_{v}$  reads
\begin{eqnarray}
a_{v}=\cos\theta_{v}+i\cos\alpha_{v}\sin\theta_{v}, \label{av}
\end{eqnarray}
 and $z^{(1)}_{{v_1}{v_2}}$, $z^{(2)}_{{v_1}{v_2}}$ are defined as
\begin{eqnarray}
z^{(1)}_{{v_1}{v_2}} =\prod_{u_1\in N_{U}(v_1)} \prod_{\substack{u_2\in N_{U}(v_2)\\ u_2 \neq u_1}}(\cos \phi_{u_1v_1} +\nonumber\\+i\sin\theta_{u_1}\cos\alpha_{u_1}\sin \phi_{u_1v_1} )\times\nonumber\\ \times (\cos \phi_{u_2v_2} +i\sin\theta_{u_2}\cos\alpha_{u_2}\sin \phi_{u_2v_2} )\times\nonumber\\ 
\prod_{w_1\in N_{W}(v_1)} \prod_{\substack{w_2\in N_{W}(v_2)\\ w_2 \neq w_1}} (\cos \phi_{v_1w_1} +i\cos\theta_{w_1}\sin \phi_{v_1w_1} )\times\nonumber\\ \times(\cos \phi_{v_2w_2} +i\cos\theta_{w_2}\sin \phi_{v_22_2} )\times\nonumber\\ 
\prod_{u_3\in N_{U}(v_1,v_2)} (\cos(\phi_{v_1u_3}+\phi_{v_2u_3}) +\nonumber\\+i\sin\theta_{u_3}\cos\alpha_{u_3}\sin (\phi_{v_1u_3}+\phi_{v_2u_3}) )\nonumber\\ \prod_{w_3\in N_{W}(v_1,v_2)}(\cos(\phi_{v_1w_3}+\phi_{v_2w_3}) +\nonumber\\+i\cos\theta_{w_3}\sin (\phi_{v_1w_3}+\phi_{v_2w_3}) ),\nonumber\\
\end{eqnarray}

\begin{eqnarray}
z^{(2)}_{{v_1}{v_2}} =\prod_{u_1\in N_{U}(v_1)} \prod_{\substack{u_2\in N_{U}(v_2)\\ u_2 \neq u_1}}(\cos \phi_{u_1v_1} +\nonumber\\+i\sin\theta_{u_1}\cos\alpha_{u_1}\sin \phi_{u_1v_1} )\times\nonumber\\ \times (\cos \phi_{u_2v_2} -i\sin\theta_{u_2}\cos\alpha_{u_2}\sin \phi_{u_2v_2} )\times\nonumber\\ 
\prod_{w_1\in N_{W}(v_1)} \prod_{\substack{w_2\in N_{W}(v_2)\\ w_2 \neq w_1}} (\cos \phi_{v_1w_1} +i\cos\theta_{w_1}\sin \phi_{v_1w_1} ) \times\nonumber\\ \times (\cos \phi_{v_2w_2} -i\cos\theta_{w_2}\sin \phi_{v_2w_2} )\times\nonumber\\ 
\prod_{u_3\in N_{U}(v_1,v_2)} (\cos(\phi_{v_1u_3}-\phi_{v_2u_3}) +\nonumber\\+i\sin\theta_{u_3}\cos\alpha_{u_3}\sin (\phi_{v_1u_3}-\phi_{v_2u_3}) )\nonumber\\ \prod_{w_3\in N_{W}(v_1,v_2)}(\cos(\phi_{v_1w_3}-\phi_{v_2w_3}) +\nonumber\\+i\cos\theta_{w_3}\sin (\phi_{v_1w_3}-\phi_{v_2w_3}) ),
\end{eqnarray}
And we can also write  $\bra{\psi} \sigma^y_{v_1}  \sigma^y_{v_2}  \ket{\psi}=\sin\theta_{v_1}\sin\theta_{v_2}\sin\alpha_{v_1}\sin\alpha_{v_2}.$ In particular case of unweighted graph  $\phi_{uv}=\phi_{uw}=\phi_{vw}=\phi$, and 
$\theta_{u_i}=\theta^{(U)}$, $\alpha_{u_i}=\alpha^{(U)}$, $\theta_{W_i}=\theta^{(W)}$, $\alpha_{w_i}=\alpha^{(W)}$, we obtain
\begin{eqnarray}
z^{(1)}_{{v_1}{v_2}} =\nonumber\\=(\cos \phi +i\sin\theta^{(U)}\cos\alpha^{(U)}\sin \phi )^{\lvert N_U(v_1) \,\triangle\, N_U(v_2) \rvert}\times\nonumber\\ \times (\cos \phi +i\cos\theta^{(W)}\sin \phi )^{\lvert N_W(v_1) \,\triangle\, N_W(v_2) \rvert}\times\nonumber\\ \times
 (\cos(2\phi) +i\sin\theta^{(U)}\cos\alpha^{(U)}\sin (2\phi) )^{\lvert N_U(v_1) \,\cap\, N_U(v_2) \rvert}\times\nonumber\\ \times(\cos(2\phi) +i\cos\theta^{(W)}\sin (2\phi) )^{\lvert N_W(v_1) \,\cap\, N_W(v_2) \rvert},\nonumber\\
\end{eqnarray} 
\begin{eqnarray}
z^{(2)}_{{v_1}{v_2}} =\nonumber\\(\cos \phi +i\sin\theta^{(U)}\cos\alpha^{(U)}\sin \phi )^{\lvert N_U(v_1)/N_U(v_2) \rvert}\times\nonumber\\ \times (\cos \phi -i\sin\theta^{(U)}\cos\alpha^{(U)}\sin \phi )^{\lvert N_U(v_2)/N_U(v_1) \rvert}\times\nonumber\\ \times
 (\cos \phi +i\cos\theta^{(W)}\sin \phi )^{\lvert N_W(v_1)/N_W(v_2) \rvert}\times\nonumber\\ \times(\cos \phi -i\cos\theta^{(W)}\sin \phi )^{\lvert N_W(v_2)/N_W(v_1) \rvert}.\nonumber\\
\end{eqnarray}

So, the quantum values $\bra{\psi} \sigma^{x}_{v_1}  \sigma^x_{v_2}  \ket{\psi}$, $\bra{\psi} \sigma^z_{v_1}  \sigma^z_{v_2}  \ket{\psi}$, $\bra{\psi} \sigma^z_{v_1}  \sigma^x_{v_2}  \ket{\psi}$  with ${v_1} \in V$, ${v_2}\in V$ depend on the structural properties of the tripartite graph. That is, they depend on the number of different neighbors (non-overlapping) of $v_1$, $v_2$,  $\lvert N_A(v_1) \,\triangle\, N_A(v_2) \rvert$ in $A$, the 
number of common neighbors of $v_1$, $v_2$ in $A$,   and also on the number of 4-cycles 
\begin{eqnarray}
n^{(A)}_{v_1,v_2}=C^2_{\lvert N_A(v_1) \,\cap\, N_A(v_2) \rvert}
\end{eqnarray} 
involving the vertices $v_1$, $v_2$, and the vertices of the set $A$. Here, $A$ denotes $U$, $W$ in the tripartite graph.   

In addition, for quantities 
 $\bra{\psi} \sigma^{x}_{w_1}  \sigma^x_{w_2}  \ket{\psi}$, $\bra{\psi} \sigma^y_{w_1}  \sigma^y_{w_2}  \ket{\psi}$, $\bra{\psi} \sigma^z_{w_1}  \sigma^z_{w_2}  \ket{\psi}$ we find
\begin{eqnarray}
\bra{\psi} \sigma^x_{w_1}  \sigma^x_{w_2}  \ket{\psi}=\frac{1}{2} \Re (z^{(1)}_{{w_1}{w_2}} a_{w_1}a_{w_2}+z^{(2)}_{{w_1}{w_2}} a_{w_1}a^*_{w_2}),\\
\bra{\psi} \sigma^y_{w_1}  \sigma^y_{w_2}  \ket{\psi}=\frac{1}{2} \Re (-z^{(1)}_{{w_1}{w_2}} a_{w_1}a_{w_2}+z^{(2)}_{{w_1}{w_2}}a_{w_1}a^*_{w_2}),\\
\bra{\psi} \sigma^x_{w_1}  \sigma^y_{w_2}  \ket{\psi}=\frac{1}{2} \Im (z^{(1)}_{{w_1}{w_2}} a_{w_1}a_{w_2}+z^{(2)}_{{w_1}{w_2}}a_{w_1}a^*_{w_2}),
\end{eqnarray}
where for simplicity of presentation, the following notations are introduced

\begin{eqnarray}
a_w=\sin \theta_{w} e^{i \alpha_{w}},
\end{eqnarray}
\begin{eqnarray}
z^{(1)}_{{w_1}{w_2}} =\prod_{u_1\in N_{U}(w_1)} \prod_{\substack{u_2\in N_{U}(w_2)\\ w_2 \neq w_1}}(\cos \phi_{w_1u_1} +\nonumber\\+i\sin\theta_{u_1}\cos\alpha_{u_1}\sin \phi_{w_1u_1} )\times\nonumber
\end{eqnarray}
\begin{eqnarray}
\times (\cos \phi_{w_2u_2} +i\sin\theta_{u_2}\cos\alpha_{u_2}\sin \phi_{w_2u_2} )\times\nonumber\\ \times
\prod_{v_1\in N_{V}(w_1)} \prod_{\substack{v_2\in N_{V}(w_2)\\ v_2 \neq v_1}} (\cos \phi_{v_1w_1} +\nonumber\\+i\sin\theta_{v_1}\sin\alpha_{v_1}\sin \phi_{v_1w_1} )\times\nonumber
\end{eqnarray}
\begin{eqnarray}
\times(\cos \phi_{v_2w_2} +i\sin\theta_{v_2}\sin\alpha_{v_2}\sin \phi_{v_2 w_2} )\times\nonumber
\end{eqnarray}
\begin{eqnarray}
\prod_{u_3\in N_{U}(w_1,w_2)} (\cos(\phi_{w_1u_3}+\phi_{w_2u_3}) +\nonumber\\+i\sin\theta_{u_3}\cos\alpha_{u_3}\sin (\phi_{w_1u_3}+\phi_{w_2u_3}) )\nonumber\\ \prod_{v_3\in N_{V}(w_1,w_2)}(\cos(\phi_{w_1v_3}+\phi_{w_2v_3}) +\nonumber\\+i\cos\theta_{v_3}\sin (\phi_{w_1v_3}+\phi_{w_2v_3}) ),
\end{eqnarray}

\begin{eqnarray}
z^{(2)}_{{w_1}{w_2}} =\prod_{u_1\in N_{U}(w_1)} \prod_{\substack{u_2\in N_{U}(w_2)\\ w_2 \neq w_1}}(\cos \phi_{w_1u_1} +\nonumber\\+i\sin\theta_{u_1}\cos\alpha_{u_1}\sin \phi_{w_1u_1} )\times\nonumber
\end{eqnarray}
\begin{eqnarray}
\times(\cos \phi_{w_2u_2} -i\sin\theta_{u_2}\cos\alpha_{u_2}\sin \phi_{w_2u_2} )\times\nonumber\\ \times
\prod_{v_1\in N_{V}(w_1)} \prod_{\substack{v_2\in N_{V}(w_2)\\ v_2 \neq v_1}} (\cos \phi_{v_1w_1} +\nonumber\\+i\sin\theta_{v_1}\sin\alpha_{v_1}\sin \phi_{v_1w_1} )\times\nonumber\\ \times(\cos \phi_{v_2w_2} -i\sin\theta_{v_2}\sin\alpha_{v_2}\sin \phi_{v_2 w_2} )\times\nonumber\\ \times
\prod_{u_3\in N_{U}(w_1,w_2)} (\cos(\phi_{w_1u_3}-\phi_{w_2u_3}) +\nonumber\\+i\sin\theta_{u_3}\cos\alpha_{u_3}\sin (\phi_{w_1u_3}+\phi_{w_2u_3}) )\nonumber\\ \prod_{v_3\in N_{V}(w_1,w_2)}(\cos(\phi_{w_1v_3}-\phi_{w_2v_3}) +\nonumber\\+i\cos\theta_{v_3}\sin (\phi_{w_1v_3}+\phi_{w_2v_3}) ).
\end{eqnarray}
We also have $\bra{\psi} \sigma^z_{w_1}  \sigma^z_{w_2}  \ket{\psi}=\cos\theta_{w_1}\cos\theta_{w_2}$. In the case of   $\phi_{uv}=\phi_{uw}=\phi_{vw}=\phi$, and also the initial states of all qubits representing one set are the same 
$\theta_{u_i}=\theta^{(U)}$, $\alpha_{u_i}=\alpha^{(U)}$, $\theta_{v_i}=\theta^{(V)}$, $\alpha_{v_i}=\alpha^{(V)}$, we can write

\begin{eqnarray}
z^{(1)}_{{w_1}{w_2}} =\nonumber\\=(\cos \phi +i\sin\theta^{(U)}\cos\alpha^{(U)}\sin \phi )^{\lvert N_U(w_1) \,\triangle\, N_U(w_2) \rvert}\times\nonumber\\ (\cos \phi +i\sin\theta^{(V)}\cos\alpha^{(V)}\sin \phi )^{\lvert N_V(w_1) \,\triangle\, N_V(w_2) \rvert}\times\nonumber\\ 
 \times (\cos(2\phi) +\nonumber\\+i\sin\theta^{(U)}\cos\alpha^{(U)}\sin (2\phi) )^{\lvert N_U(w_1) \,\cap\, N_U(w_2) \rvert}\times\nonumber\\ (\cos(2\phi) +i\cos\theta^{(V)}\sin (2\phi) )^{\lvert N_V(w_1) \,\cap\, N_V(w_2) \rvert}, 
 \end{eqnarray}
\begin{eqnarray}
z^{(2)}_{{w_1}{w_2}} =\nonumber\\=(\cos \phi +i\sin\theta^{(U)}\cos\alpha^{(U)}\sin \phi )^{\lvert N_U(w_1)/N_U(w_2) \rvert}\times\nonumber\\ \times (\cos \phi -i\sin\theta^{(U)}\cos\alpha^{(U)}\sin \phi )^{\lvert N_U(w_2)/N_U(w_1) \rvert}\times\nonumber\\ \times
 (\cos \phi +i\sin\theta^{(V)}\sin\alpha^{(V)}\sin \phi )^{\lvert N_V(w_1)/N_V(w_2) \rvert}\times\nonumber\\ \times(\cos \phi -\nonumber\\-i\sin\theta^{(V)}\sin\alpha^{(V)}\sin \phi )^{\lvert N_V(w_2)/N_V(w_1) \rvert}.
\end{eqnarray}

So, the quantum correlators 
$\bra{\psi} \sigma^{x}_{w_1} \sigma^{x}_{w_2} \ket{\psi}$, 
$\bra{\psi} \sigma^{y}_{w_1} \sigma^{y}_{w_2} \ket{\psi}$, and 
$\bra{\psi} \sigma^{x}_{w_1} \sigma^{y}_{w_2} \ket{\psi}$, 
for $w_1, w_2 \in W$, depend on the number of non-overlapping neighbors of $w_1$ and $w_2$ in $A$, given by 
$\lvert N_A(v_1) \triangle N_A(v_2) \rvert$, 
the number of their common neighbors in $A$, and the number of 4-cycles 
$n^{(A)}_{w_1,w_2}=C^2_{\lvert N_A(w_1) \,\cap\, N_A(w_2) \rvert}$.
formed by the vertices $w_1$, $w_2$, and vertices from the set $A$. Here $A$ denotes either $U$ or $V$ in the tripartite graph.

\section{Quantum calculations of the entanglement distance of tripartite quantum graph states}\label{s3}

Due to the explicit relation between the entanglement distance and the mean spin, the entanglement measure can be quantified directly using quantum computing. We now consider the particular case of a quantum graph state corresponding to a triangle and compute the entanglement of one qubit with the remaining qubits using quantum computing. The state reads

\begin{eqnarray}
\ket{\psi}=RXY_{01}(\phi_{01})RYZ_{12}(\phi_{12})\times \nonumber\\ \times RXZ_{20}(\phi_{20})\ket{\psi^{(0)}_{init}}\ket{\psi^{(1)}_{init}}\ket{\psi^{(2)}_{init}}.\label{state_tr}
\end{eqnarray}
For convenience we use the following notation for the initial states
\begin{eqnarray}
\ket{\psi^{(A)}_{init}}=\prod_{k\in A}\left(\cos \frac{\theta^{(A)}_k}{2} \ket{0}_k + e^{i\alpha^{(A)}_k} \sin \frac{\theta^{(A)}_k}{2} \ket{1}_k\right).
\end{eqnarray}
here $A=(0,1,2)$.  State (\ref{state_tr}) corresponds to the simplest tripartite graph that is triangle.

\begin{figure}[h!]
		\centering
	\includegraphics[scale=0.35]{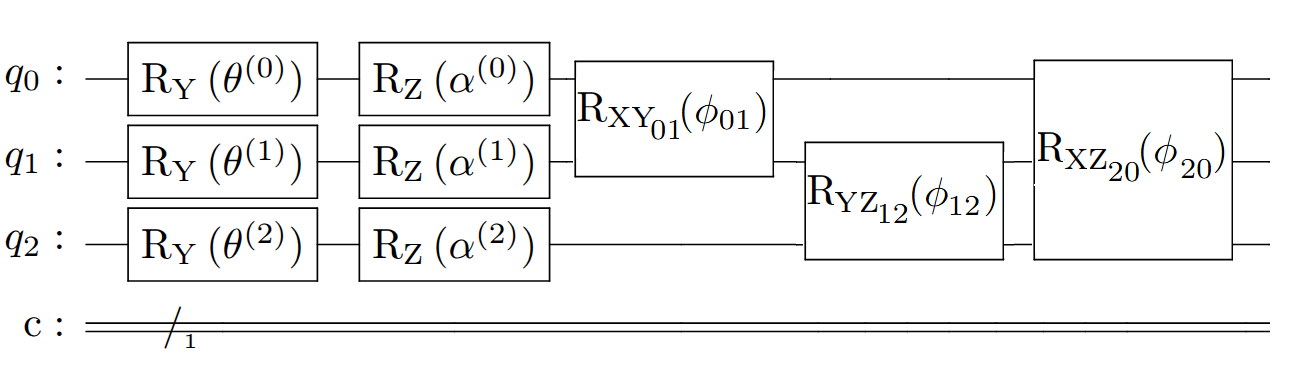}
		\caption{Quantum protocol for preparation of state  (\ref{state_tr}).} 
		\label{fig1}
\end{figure}

Quantum protocol for preparation of state (\ref{state_tr})  is presented in Fig. \ref{fig1}.

According to the definition to quantify the entanglement distance of qubit $q[k]$ with other qubits  in the graph state (\ref{state}) we have to quantify mean values of Pauli operators. To compute  $\bra{\psi} \sigma^z_k \ket{\psi}$  we prepare quantum state $\ket{\psi}$ with quantum protocol presented in Fig. \ref{fig1}. Then the mean value of $\sigma^z_k$ operator can be found on the basis of the results of measurement of state of qubit $q[k]$ in the standard basis $\langle\sigma^z_k \rangle=\vert \langle \psi \vert 0 \rangle \vert^2-\vert \langle \psi \vert 1 \rangle \vert^2$. To detect   mean value $\bra{\psi} \sigma^x_k \ket{\psi}$   we have to apply  $RY_k(-\pi/2)$ before measurement in the standard basis. To find $\bra{\psi} \sigma^y_k \ket{\psi}$ we have to apply $RX_k(\pi/2)$  before the measurement.  This is because of identities $\sigma^x_k=\exp(-i\pi\sigma^y_k/4)\sigma^z_k\exp(i\pi\sigma^y_k/4)$, $\sigma^y_k=\exp(i\pi\sigma^x_k/4)\sigma^z_k\exp(-i\pi\sigma^x_k/4)$. So, $\langle\sigma^x_k \rangle=\vert \langle \tilde\psi^{y} \vert 0 \rangle \vert^2-\vert \langle \tilde\psi^{y} \vert 1 \rangle \vert^2$, $\langle\sigma^y_k \rangle=\vert \langle \tilde\psi^{x} \vert 0 \rangle \vert^2-\vert \langle \tilde\psi^{x} \vert 1 \rangle \vert^2$,
 where $\vert\tilde\psi^{y}\rangle=RY_k(-\pi/2)\vert\psi\rangle$, $\vert\tilde\psi^{x}\rangle=RX_k(\pi/2)\vert\psi\rangle$.

\begin{figure}[h!]
		\centering
	\includegraphics[scale=0.25]{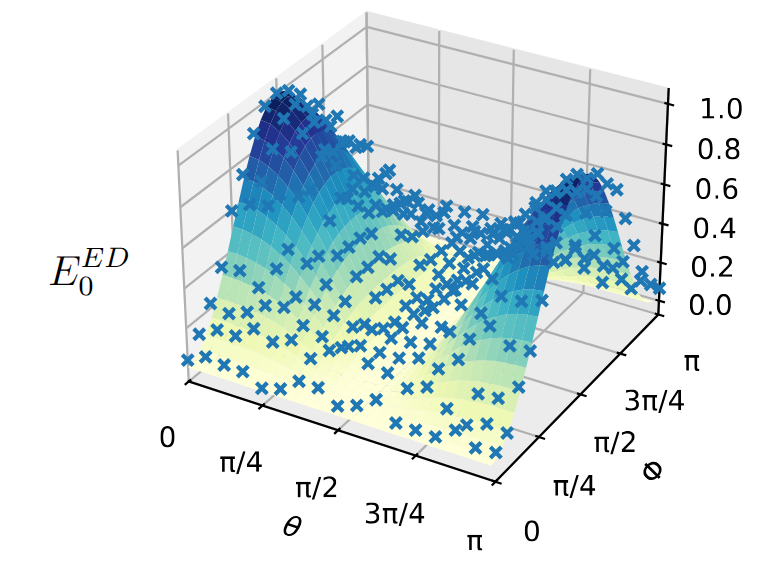}
		\caption{Entanglement distance of qubit $q[0]$ with other qubits in state (\ref{state_tr}) for $\alpha=0$ and  different values of $\theta$, $\phi$. The results obtained using the AerSimulator which includes a readout error of the order $10^{-2}$, a Pauli-X error of $10^{-4}$, and a CNOT error of $10^{-2}$ are indicated by cross markers, while the continuous surface represents the corresponding analytical calculations.}
		\label{results_entanglement0}
\end{figure}

\begin{figure}[h!]
		\centering
	\includegraphics[scale=0.3]{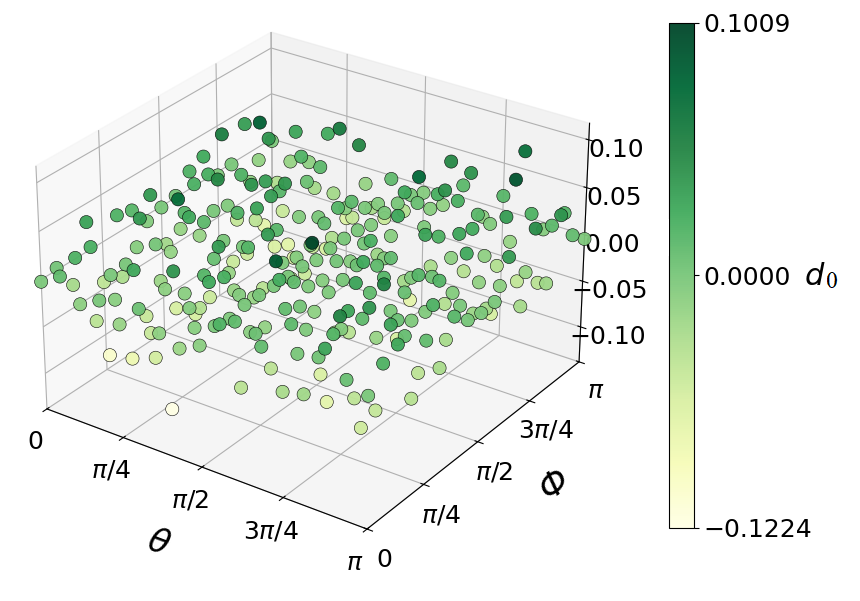}
	\caption{Differences between the analytical results for  the entanglement distance of qubit $q[0]$ with other qubits in state (\ref{state_tr}) for $\alpha=0$ and  different values of $\theta$, $\phi$. and results obtained using the AerSimulator.} 
		\label{dif0}
\end{figure}

\begin{figure}[h!]
		\centering
	\includegraphics[scale=0.25]{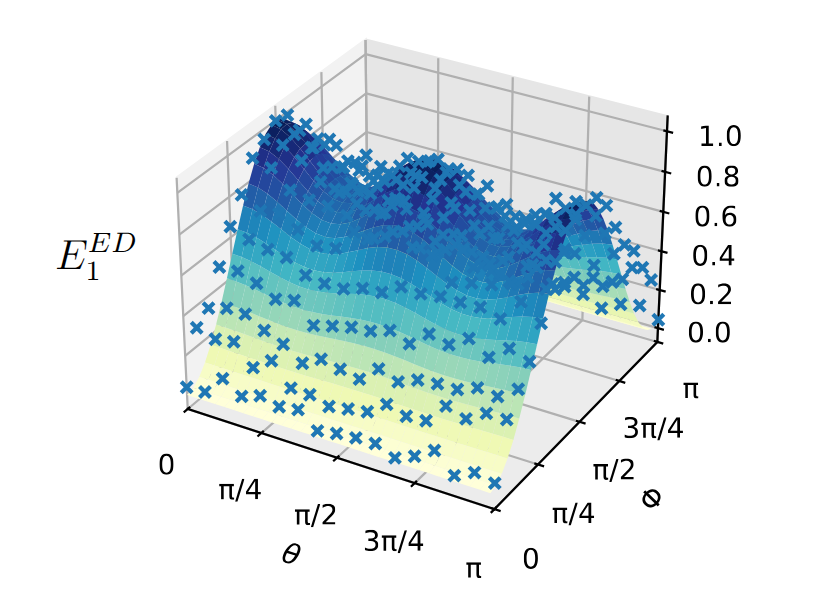}
		\caption{Entanglement distance of qubit $q[1]$ with other qubits in state (\ref{state_tr}) for $\alpha=0$ and  different values of $\theta$, $\phi$.  The results obtained using the AerSimulator which includes a readout error of the order $10^{-2}$, a Pauli-X error of $10^{-4}$, and a CNOT error of $10^{-2}$ are indicated by cross markers, while the continuous surface represents the corresponding analytical calculations.}. 
		\label{results_entanglement1}
\end{figure}

\begin{figure}[h!]
		\centering
	\includegraphics[scale=0.3]{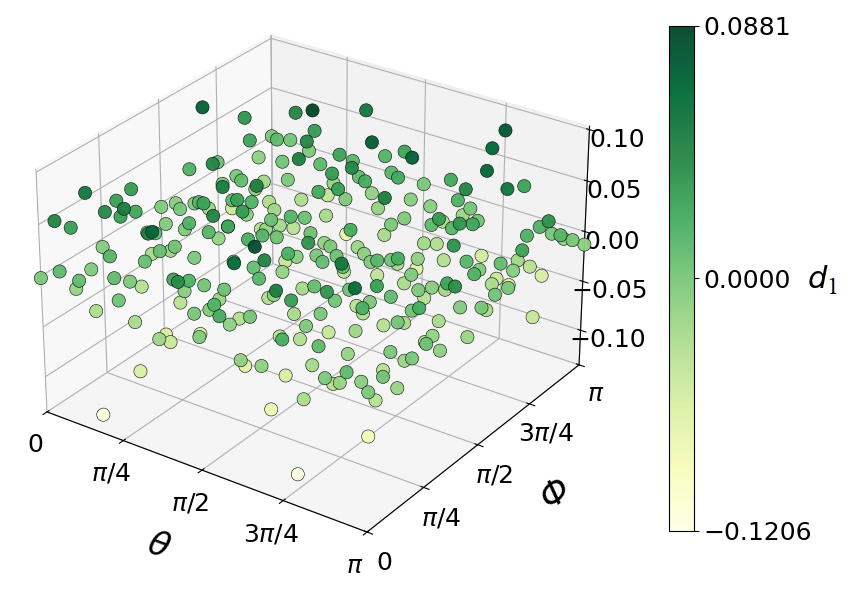}
		\caption{Differences between the analytical results for  the entanglement distance of qubit $q[1]$ with other qubits in state (\ref{state_tr}) for $\alpha=0$ and  different values of $\theta$, $\phi$. and results obtained using the AerSimulator.}
		\label{dif1}
\end{figure}

\begin{figure}[h!]
		\centering
	\includegraphics[scale=0.25]{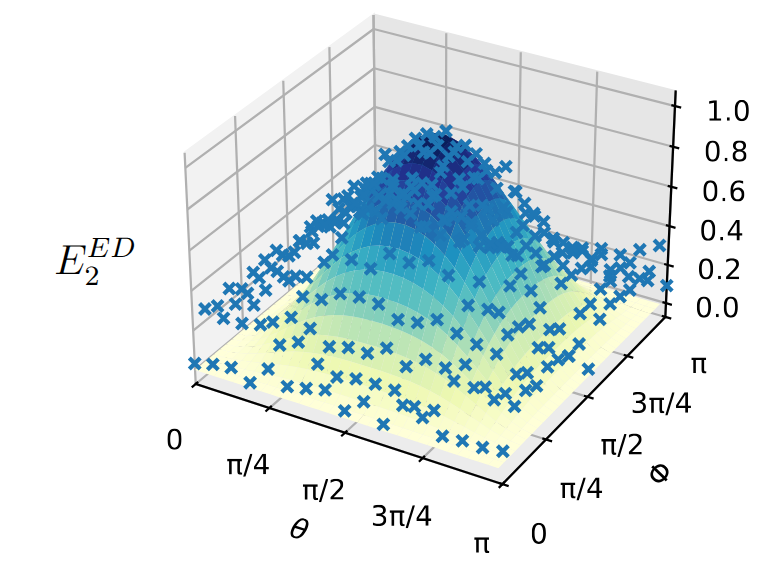}
		\caption{ Entanglement distance of qubit $q[2]$ with other qubits in state (\ref{state_tr}) for $\alpha=0$ and  different values of $\theta$, $\phi$. The results obtained using the AerSimulator which includes a readout error of the order $10^{-2}$, a Pauli-X error of $10^{-4}$, and a CNOT error of $10^{-2}$ are indicated by cross markers, while the continuous surface represents the corresponding analytical calculations.}. 
		\label{results_entanglement2}
\end{figure}

We have computed the entanglement distance of the qubits $q[k]$, with $k=(0,1,2)$, with the remaining qubits in the state (\ref{state_tr}) for $\alpha=0$ and various values of $\theta$ and $\phi$ with quantum computing, using quantum protocol Fig. \ref{fig1} and measuring mean values of corresponding Pauli operators. The numerical results obtained on the AerSimulator, which incorporates a readout error of order $10^{-2}$, a Pauli-X error of $10^{-4}$, and a CNOT error of $10^{-2}$, are shown as cross markers in Figs.~\ref{results_entanglement0}–\ref{results_entanglement2}. The corresponding analytical results are represented by the continuous surfaces in Figs.~\ref{results_entanglement0}–\ref{results_entanglement2}.  The errors in the quantum calculations are chosen to be of the same order as those of IBM quantum computers.

\begin{figure}[h!]
		\centering
	\includegraphics[scale=0.3]{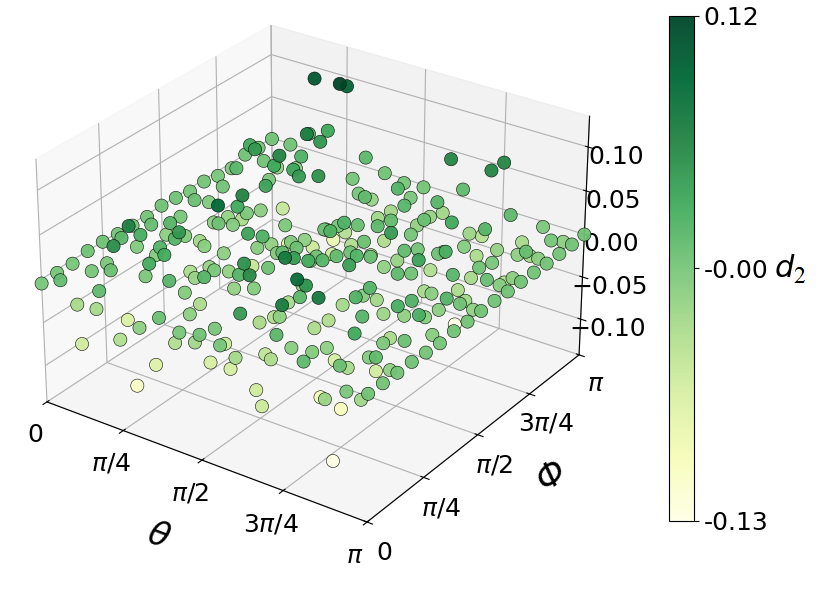}
		\caption{Differences between the analytical results for  the entanglement distance of qubit $q[2]$ with other qubits in state (\ref{state_tr}) for $\alpha=0$ and  different values of $\theta$, $\phi$. and results obtained using the AerSimulator.}
		\label{dif2}
\end{figure}

We have also computed the differences $d_k$ between the theoretical values of the entanglement distance of the qubits $q[k]$ (with $k=0,1,2$) in the state (\ref{state_tr}) for $\alpha=0$ and various values of $\theta$ and $\phi$, and the corresponding results obtained via quantum simulation on the AerSimulator.  The deviation is defined as
\begin{eqnarray}
d_k =  E^{(ED)}_k(\ket{\psi})_{\mathrm{theor}} - E^{(ED)}_k(\ket{\psi})_{\mathrm{comp}}.
\end{eqnarray}
The results are presented in Figs.~\ref{dif0}–\ref{dif2}. It is worth noting that the numerical results obtained from quantum simulations are in good agreement with the theoretical predictions.

\section{Conclusions}\label{s4}

We propose a method for constructing multi-qubit entangled quantum states that represent tripartite graphs. These states are generated by applying two-qubit gates $RXY_{kl}(\phi_{kl})$, $RXZ_{kl}(\phi_{kl})$, and $RYZ_{kl}(\phi_{kl})$ to a general separable quantum state of $n$ qubits (\ref{state}). The resulting states are entangled and can be interpreted as quantum graph states corresponding to weighted tripartite graphs.

In the general case of quantum states associated with weighted tripartite graphs of arbitrary structure the expression for the entanglement distance was obtained (14), (17), (20). It is shown that the entanglement distance  $E^{ED}_k(\ket{\psi})$ is determined by the weights of the arcs connected to the vertex $k$, characterized by the parameters $\phi_{kl}$, as well as by the parameters defining the initial states of the qubits in the closed neighborhood of $k$ within the graph $G(U,V,W,E)$ (14), (17), (20). 

In the particular case where all initial state parameters are identical and the edge weights between sets are equal, the entanglement of a given qubit with the rest of the system depends on the number of its connections to vertices in other sets,  namely it depends on the vertex  degree with respect to the sets in tripartite graph (21)-(23). 
Thus, we have derived a relation between the entanglement of a qubit and the remaining qubits in a multi-qubit quantum state representing tripartite graphs, linking it to the properties of the corresponding vertices in the graph. These results open up the possibility of studying vertex degrees with respect to different sets in a graph using a quantum computer.
As a specific example, we consider the case of a tripartite graph forming a triangle. For this system, the entanglement distance is evaluated using quantum computing on the AerSimulator, including noise models Fig. 2, Fig. 4, Fig. 6. The results of quantum computing are in good agreement with the theoretical predictions (see Fig. 3, Fig. 5, Fig. 7). These results also confirm the correctness of the derived analytical expressions. The three-qubit triangle graph state can be considered a benchmark example for validating the analytical results.

We also calculate quantum correlators for  quantum states corresponding to general tripartite graphs (see Sec. \ref{s2}). We establish a connection between these correlators and the structural properties of the underlying graphs. Namely we have shown that the correlators depend on the number of non-overlapping neighbors of the corresponding vertices, the number of their common neighbors, and the number of 4-cycles formed by these vertices together with vertices from other sets.

At the end of this section it is worth noting that tripartite graphs is an important class of multipartite structures for studying interactions among three groups of subsystems. From the perspective of quantum information, tripartite quantum graph states provide a convenient setting for investigating how the topology and weights of a graph are reflected in the entanglement properties of the corresponding quantum state. We have also found that quantum correlators are related to structural features of tripartite graphs. This establishes a connection between graph properties and characteristics of quantum states, which can be explored using quantum computing. Also, it is worth adding that the tripartite case provides a useful intermediate step between bipartite and more general multipartite graph structures and offers a tractable framework for developing and testing methods that can subsequently be extended to more general multipartite structures. 

The tripartite graphs have broad applications in various practical problems, including resource allocation, scheduling, and optimization tasks. The obtained results open up the possibility of investigating structural properties of such classical  objects using quantum computing.


\begin{thebibliography}{00}

\bibitem{Horodecki} R. Horodecki, P. Horodecki, M. Horodecki, and K. Horodecki, “Quantum entanglement,” Rev. Mod. Phys., vol. 81, pp. 865–942, 2009.



\bibitem{Feynman} R. P. Feynman, "Simulating physics with computers" Int. J. Theor. Phys. vol. 21, p. 467, 1982.
\bibitem{Ekert}   A. K. Ekert, “Quantum cryptography based on Bell’s theorem,” Phys. Rev. Lett.,
vol. 67, pp. 661–663,  1991.
\bibitem{Bennett} C. H. Bennett, G. Brassard, C. Cr{\'e}peau, R. Jozsa, A. Peres, and W. K. Wootters, “Teleporting an unknown
quantum state via dual classical and einstein-podolsky-rosen channels,” Phys. Rev. Lett., vol. 70, pp. 1895–1899, 1993.
\bibitem{Lloyd} S. Lloyd, “Universal quantum simulators,” Science, vol. 273, no. 5278, pp. 1073–1078, 1996.
\bibitem{Bouwmeester}  D. Bouwmeester, J.-W. Pan, K. Mattle, M. Eibl, H. Weinfurter, and A. Zeilinger, “Experimental quantum
teleportation,” Nature, vol. 390, pp. 575–579, 1997.


\bibitem{Raussendorf}  R. Raussendorf and H. J. Briegel, “A one-way quantum computer,” Phys. Rev. Lett., vol. 86, 
pp. 5188–5191, 2001.
\bibitem{Buluta}  I. Buluta and F. Nori, “Quantum simulators,” Science, vol. 326,  pp. 108–111, 2009.

\bibitem{Shimony}  A. Shimony, “Degree of entanglement,” Annals of the New York Academy of Sciences,
vol. 755, pp. 675–679, 1995.






\bibitem{Horodecki1}  P. Horodecki and A. Ekert, “Method for direct detection of quantum entanglement,” Phys. Rev. Lett.,
vol. 89,  p. 127902, 2002.

\bibitem{Arrigo} A. D’Arrigo, R. L. Franco, G. Benenti, E. Paladino, and G. Falci, “Recovering entanglement by local
operations,” Annals of Physics, vol. 350, pp. 211–224, 2014.

\bibitem{Shi} Sh. Shi, L. Tian, Y. Wang, Y. Zheng, Ch. Xie, and K. Peng, “Demonstration of channel multiplexing quantum communication exploiting entangled sideband modes,” Phys. Rev. Lett., vol. 125, p. 070502, 2020.

\bibitem{Llewellyn}  D. Llewellyn, Y. Ding, I. I. Faruque, S. Paesani, D. Bacco, R. Santagati, Y.-J. Qian, Y. Li, Y.-F. Xiao, M. Huber, M. Malik, G. F. Sinclair, X. Zhou, K. Rottwitt, J. L. O’Brien, J. G. Rarity, Q. Gong, L. K. Oxenlowe, J. Wang, and M. G. Thompson, “Chip-to-chip quantum teleportation and multi-photon entan-
glement in silicon,” Nature Physics, vol. 16, pp. 148–153, 2020.

\bibitem{Bell}   B. A. Bell, D. A. Herrera-Marti, and M. S. Tame, “Experimental demonstration of graph-state quantum
secret sharing,” Nature Communications, vol. 5, p. 3658, 2014.
\bibitem{Vesperini1} A. Vesperini and R. Franzosi, “Entanglement, quantum correlators, and connectivity in graph states,” Advanced
Quantum Technologies, vol. 7, p. 2300264, 2024.
\bibitem{Vesperini2} A. Vesperini, “Correlations and projective measurements in maximally entangled multipartite states,” Annals of Physics, vol. 457, p. 169406, 2023.


\bibitem{Markham} D. Markham and B. C. Sanders, “Graph states for quantum secret sharing,” Phys. Rev. A, vol. 78,
no. 4, p. 042309, 2008.
\bibitem{Wang}  Y. Wang, Y. Li, Z.-Q. Yin, and B. Zeng, “16-qubit ibm universal quantum computer can be fully entangled,”
npj Quantum Information, vol. 4, p. 46, 2018.
\bibitem{Mooney}  G. J. Mooney, C. D. Hill, and L. C. L. Hollenberg, “Entanglement in a 20-qubit
superconducting quantum
computer,”
Scientific Reports, vol. 9, p. 13465, 2019.
\bibitem{Schlingemann}  D. Schlingemann, and R. F. Werner, “Quantum error-correcting codes associated with graphs,” Phys.
Rev. A, vol. 65, p. 012308, 2001.

\bibitem{Mazurek} P. Mazurek, M. Farkas, A. Grudka, M. Horodecki, and M. Studzinski, “Quantum error-correction codes and absolutely maximally entangled states,”
Phys. Rev. A, vol. 101,  p. 042305, 2020.
\bibitem{Shettell}  N. Shettell and D. Markham, “Graph states as a resource for quantum metrology,” Phys. Rev. Lett., vol. 124, p. 110502, 2020.

\bibitem{Gnatenko24} 
Kh. P. Gnatenko, “Entanglement of multi-qubit states representing directed networks
and its detection with quantum computing,” Phys. Lett. A, vol. 521, p. 129815, 2024.


\bibitem{su} N. A. Susulovska, and  Kh. P. Gnatenko, "Quantifying geometric measure of entanglement of multi-qubit graph states on the IBM's quantum computer"
Proceedings - 2021 IEEE International Conference on Quantum Computing and Engineering, QCE 2021, 465-466, 2021.



\bibitem{Hein} M. Hein, J. Eisert, and H. J. Briegel, “Multiparty entanglement in graph states,” Phys. Rev. A,
vol. 69, p. 062311, 2004.
\bibitem{Guhne} O. G\"uhne, G. T\'oth,  P. Hyllus, and H. J. Briegel, “Bell inequalities for graph states,” Phys. Rev.
Lett., vol. 95,  p. 120405, 2005.
\bibitem{Qian}  Y. Qian, Z. Shen, G. He, and G. Zeng, “Quantum-cryptography network via continuous-variable graph states,” Phys. Rev. A, vol. 86,
 p. 052333, 2012.


\bibitem{Mezher} R. Mezher, J. Ghalbouni, J. Dgheim, and D. Markham, “Efficient quantum pseudorandomness with simple graph states,” Phys.
Rev. A, vol. 97,  p. 022333, 2018.
\bibitem{Gnatenko2026} Kh. P. Gnatenko, “Studies of properties of bipartite graphs with quantum programming,” Phys. Lett. A, vol. 566, p. 131191, 2026.


\bibitem{Susulovska} Kh. P. Gnatenko and N. A.  Susulovska, "Geometric measure of entanglement of multi-qubit graph states and its detection on a quantum computer," Europhys. Lett. vol. 136, p. 40003, 2021.

\bibitem{Vesperini} A. Vesperini, G. Bel-Hadj-Aissa, L. Capra, and R. Franzosi, “Unveiling the geometric meaning of quantum entanglement: Discrete and continuous variable systems,” Front. Phys., vol. 19, p. 51204, 2024.

\bibitem{Gao} X. Gao, Z.-Y. Zhang, and L.-M. Duan, “An efficient quantum algorithm for generative machine learning,”
Science Advances, vol. 4, no. 12, p. eaat9004, 2018.
\bibitem{Zoufal}  C. Zoufal, A. Lucchi, and S. Woerner, “Quantum generative adversarial networks for learning and loading
random distributions,” npj Quantum Information, vol. 5, p. 103, 2019.

\bibitem{Cocc}  D. Cocchiarella, S. Scali, B. N. Ribisi, G. Bel-Hadj-Aissa, and R. Franzosi, “Entanglement distance for arbitrary M-qudit hybrid systems” Phys. Rev. A, vol. 101, no. 4, p. 042129, 2020.
\bibitem{GnatenkoIEEE} Kh. P. Gnatenko, "Quantum and classical calculations of the entanglement distance of multi-qubit one-layer variational quantum states"
2025 IEEE International Conference on Quantum Computing and Engineering (QCE), Albuquerque, NM, USA, 2025, 470–471, 2025.





\bibitem{ibm} IBM Quantum Documentation.\\
https://quantum.cloud.ibm.com/docs/en/guides/local-simulators
\end{thebibliography}
\end{document}